\documentclass[numberedappendix]{./emulateapj}
\usepackage{amssymb,amsmath,amsthm}
\usepackage{longtable}
\renewcommand{\thefootnote}{\arabic{footnote}}
\def\he {\texttt{He  }}

\def\chandra{\emph{Chandra}}
\def\apj{\texttt{ApJ}}

\def\xmm{\emph{XMM-Newton}}

\begin{document}


\title{An Analytic Model of the Physical Properties of
Galaxy Clusters }


\author{
G.~Esra~Bulbul\altaffilmark{1},
Nicole~Hasler\altaffilmark{1},
Massimiliano~Bonamente\altaffilmark{1,2}
and Marshall~Joy\altaffilmark{2}}

\altaffiltext{1}{Department of Physics, University of Alabama, Huntsville, AL 35899}
\altaffiltext{2}{Space Science Office, VP62, NASA Marshall Space Flight Center, Huntsville, AL 35812}

\begin{abstract}
We introduce an analytic model of the diffuse intergalactic medium
in galaxy clusters based on a polytropic equation of state
for the gas in hydrostatic equilibrium with the cluster
gravitational potential.
This model is directly applicable to the analysis of X-ray and Sunyaev-Zeldovich
Effect observations from the cluster core to the virial radius,
with 5 global parameters and 3 parameters describing the cluster core.
We validate the model using \chandra\ X-ray observations of two
polytropic clusters, MS~1137.5+6625 and CL~J1226.9+3332,
and two cool core clusters, Abell~1835 and Abell~2204. We show that the model
accurately describes the spatially resolved spectroscopic and imaging data,
including the cluster core region where significant cooling of
the plasma is observed.
\end{abstract}
\section{Introduction}

Galaxy cluster masses play an important role in addressing
fundamental physical and cosmological problems,
such as the measurement of the cluster gas mass fraction \citep[][]{allen2008,ettori2009},
the evolution of the growth of structure
\citep[][]{mantz2008,mantz2009,vikhlinin2009},  and 
the gravitational sedimentation of ions \citep{chuzhoy2003,peng2009,shtykovskiy2010}.
A vital tool for the measurement of cluster masses is the
diffuse hot intergalactic medium 
which can be detected primarily through its bright X-ray emission
\citep[][]{sarazin1988}, or through the Sunyaev-Zeldovich Effect
 \citep[SZE,][]{sunyaev1972, carlstrom2002}. 
A variety of models are used to describe the distribution of
the gas,
from the simple isothermal $\beta$ model \citep[][]{cavaliere1976,birkinshaw1991}
to more complex models that describe either the 
X-ray properties \citep[gas density and temperature,][]{vikhlinin2006,cavaliere2009} or the 
SZE properties \citep[gas pressure,][]{nagai2007,mroczkowski2009,arnaud2009}.

We investigate a model of galaxy clusters based on 
an analytic distribution for the cluster mass density
inspired by the \citet[][]{navarro1996} distribution,
which we generalize following 
\citet{suto1998}, \citet{ascasibar2003} and \citet{ascasibar2008} to include
a variable asymptotic slope at large radii. 
This mass density is combined with a
polytropic equation of state for the gas,
to provide  self-consistent  density, temperature and pressure profiles 
for a plasma in hydrostatic equilibrium. 
The use of a polytropic equation of state for the cluster gas has also been recently
proposed by \citet{ascasibar2008} and \citet{bode2009}, and tested
observationally by \citet{sanderson2009}.

In this paper we derive analytic radial
profiles for the physical quantities (temperature, density and pressure)
relevant to X-ray and SZE observations,
and present an application of these models to
high resolution \chandra\ observations of the galaxy clusters
MS~1137.5+6625, CL~J1226.9+3332, Abell~1835 and Abell~2204.
Applications of this new model include measurement of gas mass
fraction from joint X-ray and Sunyaev-Zeldovich Effect 
observations \citep{hasler2010} and the effect of \he\ sedimentation 
on X-ray mass estimates \citep{bulbul2010}.
This paper is organized as follows: in \S \ref{sec_models}
we describe our model,
in \S \ref{sec_dataAnalysis} we present the application of the model
to \chandra\ X-ray observations of MS~1137.5+6625, CL~J1226.9+3332, 
Abell~2204 and Abell~1835, and in \S  \ref{sec:comparison} we perform a comparison between our
mass measurements and the results of \citet{mroczkowski2009}. In 
\S \ref{sec_conclusion} we
present our conclusions. 
In the analysis of the \chandra\ data  we assume the cosmological parameters
$h=0.73$, $\Omega_M=0.27$ and $\Omega_{\Lambda}=0.73$.

\section{A Model of the Intergalactic Medium Based on Hydrostatic Equilibrium and 
the  Polytropic Equation of State}
\label{sec_models}

\subsection{The Mass Density Distribution}
\label{sec_models_darkmatterProf}

The cluster gravitational potential is dominated by
dark matter, 
with the intergalactic medium and stars 
contributing less than approximately $\sim$20\% 
of the mass \citep[][]{allen2004,
allen2008,ettori2009,vikhlinin2009}.
We therefore start with a total mass density distribution
that is obtained as a generalization of the \citet{navarro1996}
profile:

\begin{equation}
\rho_{tot}(r)= \frac{\rho_{i}}{(r/r_{s})(1+r/r_{s})^{\beta}}
\label{eqn_gen_nfw_density}
\end{equation}

where $\rho_{i}$ is the normalization constant, $r_{s}$ is 
a characteristic scale radius and $\beta$+1 is the slope 
of the density  distribution at large radii. Equation \ref{eqn_gen_nfw_density}
is a simplified version of the density distribution 
introduced by \citet{suto1998}.

The total mass enclosed within radius \textit{r} can be found by taking the volume 
integral of the density (Equation~\ref{eqn_gen_nfw_density}):

\begin{equation}
M(r)=\frac{4\pi\rho_{i}r_{s}^{3}}{(\beta-2)}\left( \frac{1}{\beta-1} +\frac{1/(1-\beta) - r/r_s}{(1+r/r_{s})^{\beta-1}}\right).
\label{eqn_gen_nfw_totalMass}
\end{equation}

Equation \ref{eqn_gen_nfw_totalMass} is indeterminate at 
$\beta=2$; the limit at $\beta=2$ can be determined using L'Hospital's rule:

\begin{equation}
 M(r) = 4\pi\rho_{i}r_{s}^{3} \left( \ln(1+r/r_s) - \frac{r/r_s}{1+r/r_s}\right)\ \ [\beta=2]
\label{eqn_totalMass_at_beta2}
\end{equation}

and therefore the mass is a continuous function of $\beta$ with no discontinuity
at $\beta=2$.

The gravitational potential at a distance \textit{r} is found by 

\begin{equation}
d\phi(r) = G M(r)/r^2 dr
\label{eqn_potential_at_infinity}
\end{equation}

using the boundary condition  $\phi(\infty)=0$. Equation
\ref{eqn_potential_at_infinity} can be integrated analytically

\begin{equation}
\phi(r) = \phi_{0} \left(\frac{1}{(\beta-2)}\frac{(1+r/r_{s})^{\beta-2}-1}{r/r_{s}(1+r/r_{s})^{\beta-2}} \right),
\label{eqn_generalized_nfw_potential}
\end{equation}

where 
\begin{equation}
\phi_{0}= -\frac{4\pi G \rho_{i}r_{s}^{2}}{(\beta-1)}.
\label{eqn_phi0}
\end{equation}

The potential at $\beta=2$ is also found by using L'Hospital's rule,

\begin{equation}
\phi(r) = \phi_0 \left( \frac{\ln(1+r/r_s)}{r/r_s} \right) \ \ [\beta=2].
\end{equation}

\begin{figure}[!h]
\centering
\includegraphics[angle=-90,width=7cm]{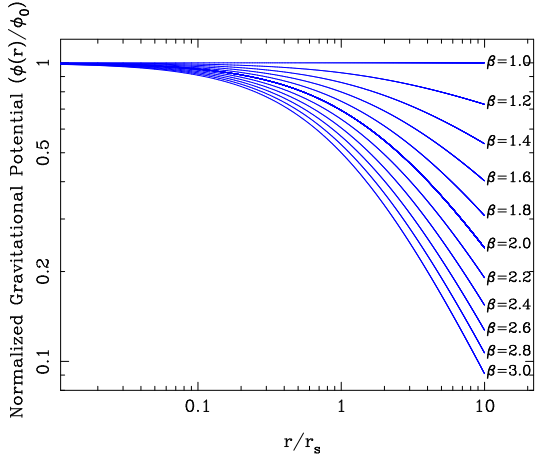}
\caption{Normalized gravitational potential for various values of
the $\beta$ parameter.}
\label{fig:gravPoten_genNFW}
\end{figure}

Therefore the gravitational potential is a continuous function of $\beta$
with no discontinuity at $\beta=2$.
Figure \ref{fig:gravPoten_genNFW} shows 
the radial distribution of the gravitational potential
for $1.0\leq \beta \leq3.0$.
The limiting value of $\beta=1$ is shown 
in Figure~\ref{fig:gravPoten_genNFW}, corresponding to
a constant potential.

\subsection{Gas Density and Temperature Profile}
\label{sec_models_gasDensityProf}

The diffuse gas is assumed to be in
 hydrostatic equilibrium with the gravitational
potential. Assuming spherical symmetry, 

\begin{equation}
\frac{1}{\mu m_p n_{e}(r)}\frac{dP_{e}}{dr} = -\frac{d\phi(r)}{dr}
\label{eqn_hydrostatic}
\end{equation}

where $P_e(r) = n_{e}(r)k T(r)$ is the electron pressure, 
\textit{G}  denotes the gravitational 
constant, $m_{p}$ is the proton mass, $\mu$ is the mean molecular weight 
of the plasma, \textit{k} is the 
Boltzmann constant and $n_{e}(r)$ is the electron number density.
In order to solve Equation~\ref{eqn_hydrostatic}, 
we assume that the gas follows a  polytropic equation of state,

\begin{equation}
\frac{n_{e,poly}(r)}{n_{e0}}=\left[\frac{T_{poly}(r)}{T_{0}} \right]^{n}
\label{eqn_polytropic}
\end{equation}

where \textit{n} is the polytropic index, $n_{e0}$ and $T_{0}$ are the values
of the  number density and temperature at $r=0$.
The polytropic index $n>0$ is a free parameter of the model, 
with the limit $n \to \infty$ describing an isothermal
distribution of gas \citep{eddington1926}.

The  temperature profile is obtained as a function
of the gravitational potential from Equations~\ref{eqn_hydrostatic} and~\ref{eqn_polytropic},

\begin{equation}
T_{poly}(r) = -\frac{1}{(n+1)}\frac{\mu m_{p}}{k} \phi(r)
\label{eqn_polytropic_hydrostatic_temperature}
\end{equation}

and therefore, using Equation \ref{eqn_generalized_nfw_potential},

\begin{equation}
T_{poly}(r)=T_{0}\left(\frac{1}{(\beta-2)}\frac{(1+r/r_{s})^{\beta-2}-1}{r/r_{s}(1+r/r_{s})^{\beta-2}} \right),
\label{eqn:polytropic_temperature}
\end{equation} 

where the normalization constant $T_{0}$ is obtained from Equations \ref{eqn_phi0}
and \ref{eqn_polytropic_hydrostatic_temperature}:

\begin{equation}
T_{0}=\frac{4\pi G \mu m_{p}}{k(n+1)}\frac{r_{s}^{2}\rho_{i}}{(\beta-1)}.
\label{eqn_relationof_T_n}
\end{equation}

Equation \ref{eqn_polytropic_hydrostatic_temperature} shows that $T_{poly}(r)\propto \phi(r)$,
and therefore Figure~\ref{fig:gravPoten_genNFW} also describes $T_{poly}(r)$ as function of radius.
Equation \ref{eqn_relationof_T_n} links the gas temperature to the normalization
of
the matter density $\rho_i$, and therefore the depth of the gravitational  potential can be determined
from the observed temperature profile.

Using the relation between temperature and gas density \
provided by the polytropic relation (Equation \ref{eqn_polytropic}), 
the polytropic gas density profile is

\begin{equation}
n_{e,poly}(r)= n_{e0} 
\left(\frac{1}{(\beta-2)}\frac{(1+r/r_{s})^{\beta-2}-1}{r/r_{s}(1+r/r_{s})^{\beta-2}} \right)^{n}.
\label{eqn:ne0_poly}
\end{equation}


\subsection{The Electron Gas Pressure}

The gas pressure is obtained using the ideal gas law $P_{e}(r)= n_{e,poly} k T_{poly}(r)$,

\begin{equation}
P_{e}(r)= P_{e0} \left(\frac{1}{(\beta-2)}\frac{(1+r/r_{s})^{\beta-2}-1}{r/r_{s}(1+r/r_{s})^{\beta-2}} \right)^{n+1}
\label{eqn:pressure}
\end{equation}

In the limit $\beta \to 2$ the pressure is analytically described by

\begin{equation}
P_{e}(r)=  P_{e0} \left(\frac{\ln(1+r/r_s)}{r/r_s}\right)^{n+1} \hspace{0.5cm} [\beta=2].
\end{equation}

The electron pressure for this model has only 4 free parameters,
and it is suitable for the analysis of Sunyaev-Zeldovich Effect observations of
galaxy clusters \citep{hasler2010}.

\subsection{Cool Core Clusters}
\label{sec_models_coolcore}

\begin{figure}[!h]
\centering
\includegraphics[angle=-90,width=7cm]{f2a.eps}
\includegraphics[angle=-90,width=7cm]{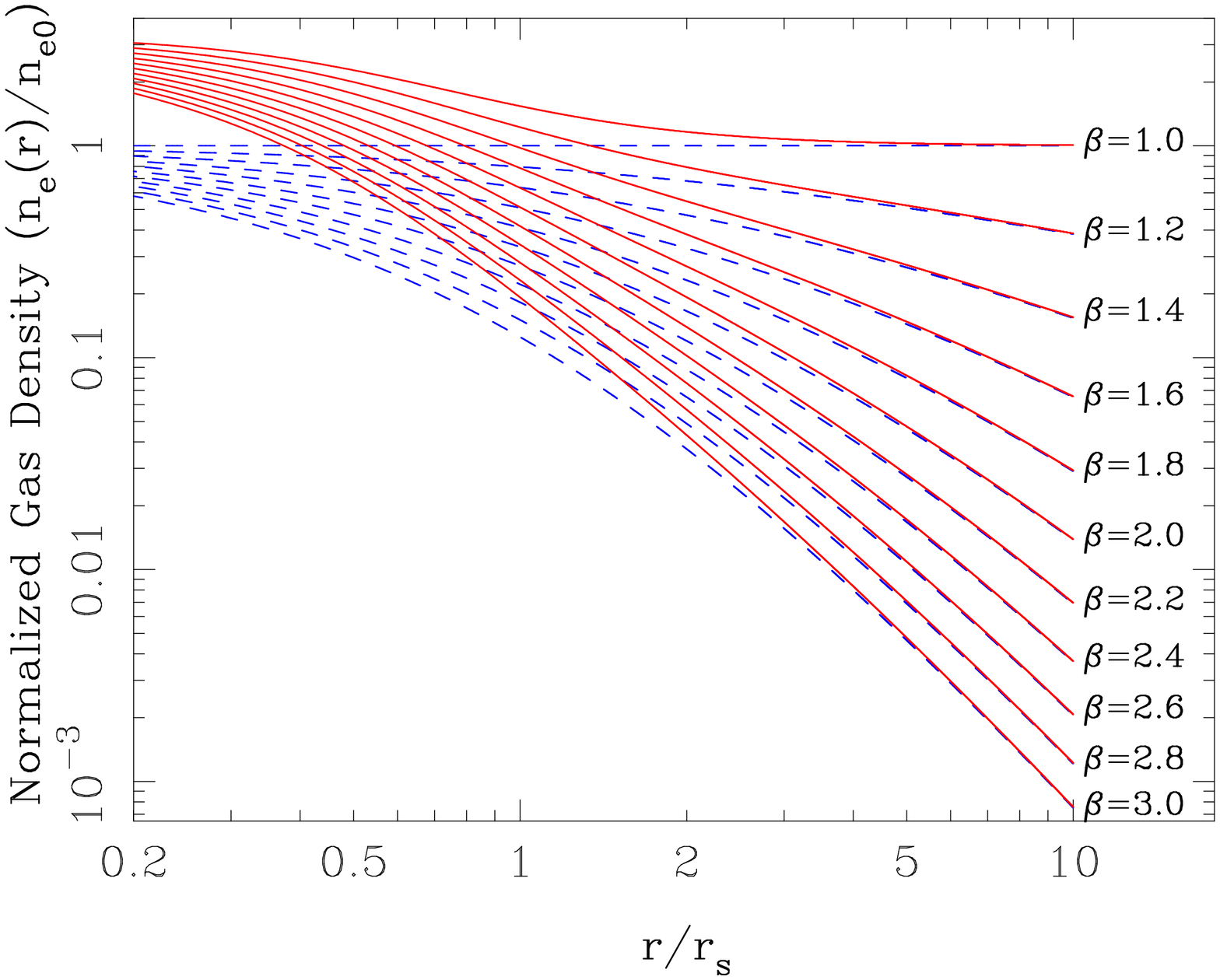}
\caption{Solid red lines are the normalized temperature and density profiles
for a taper function with parameters $\alpha=0.3$, $r_{cool}=r_{s}$ and $\gamma=2.0$, 
and variable values for $\beta$;
blue dashed lines are the models without the core taper function.}
\label{fig:tnProf}
\end{figure}

Although the temperature profile predicted by the polytropic 
model provides a good description at 
intermediate to large radii, cool core clusters feature 
a significant temperature drop in the central region
which cannot be approximated by a polytropic equation of state
\citep[see for example,][]{vikhlinin2005,sanderson2006,vikhlinin2006,baldi2007}. For cool core 
clusters  we introduce a modified  temperature profile 

\begin{equation}
\begin{aligned}
T(r)=& T_{poly}(r) \tau_{cool}(r) \\
\end{aligned}
\label{eqn_coolCore_tempProfDefn}
\end{equation}

where $T_{poly}(r)$ is the temperature profile according to the
polytropic equation of state (Equation~\ref{eqn:polytropic_temperature})
and  $\tau_{cool}(r)$ is a  phenomenological  core taper function
used by \citet{vikhlinin2006}:

\begin{equation}
\tau_{cool}(r)=\frac{\alpha+(r/r_{cool})^{\gamma}}{1+(r/r_{cool})^{\gamma}},
\label{eqn_cooling_funct}
\end{equation}

where $0 < \alpha < 1$ is a free parameter that measures the amount of central cooling and $r_{cool}$
is a characteristic cooling radius. The temperature profile modified by the
core taper function is shown in Figure \ref{fig:tnProf}
for representative values of parameters $r_{cool}$, $\gamma$ and $\alpha$.

Therefore, the explicit temperature profile for cool core clusters
is given by

\begin{equation}
T(r)=T_{0}\left(\frac{1}{(\beta-2)}\frac{(1+r/r_{s})^{\beta-2}-1}{r/r_{s}(1+r/r_{s})^{\beta-2}} \right)\tau_{cool}(r).
\label{eqn_coolCore_tempProf}
\end{equation}

In order to calculate the density distribution for cool core clusters,
we assume that the pressure distribution is the same as in the polytropic case 
(Equation \ref{eqn:pressure}). Therefore, the electron density is given by

\begin{equation}
\begin{aligned}
n_{e}(r)=&\frac{P_{e}(r)}{k T(r)}\\
=& n_{e0} \left(\frac{1}{(\beta-2)}\frac{(1+r/r_{s})
^{\beta-2}-1}{r/r_{s}(1+r/r_{s})^{\beta-2}} \right)^{n}\tau_{cool}^{-1}(r)\\
\end{aligned}
\label{eqn_coolcore_gas_density}
\end{equation}

The behavior of the gas density 
for various core taper parameters is shown in Figure \ref{fig:tnProf}.

For hydrostatic equilibrium to be satisfied, these modified 
density and temperature distributions require a modified 
total mass distribution:

\begin{equation}
\begin{aligned}
&M(r)=\\
&\frac{4\pi\rho_{i}r_{s}^{3}}{(\beta-2)}\left( \frac{1}{\beta-1} +\frac{1/(1-\beta) - r/r_s}{(1+r/r_{s})^{\beta-1}}\right)\tau_{cool}(r).
\end{aligned}
\label{eqn:gen_nfw_totalMass2}
\end{equation}

The only difference between the cool core total mass distribution (Equation 
\ref{eqn:gen_nfw_totalMass2}) and the polytropic total mass distribution (Equation 
\ref{eqn_gen_nfw_totalMass}) is the term $\tau_{cool}(r)$, which is significant
only at small radii. At large radii , the effect of the core taper vanishes, and
the thermodynamics of the gas is described by the polytropic equation of state.

\begin{table*}
\caption{Cluster Sample}
\centering
\begin{tabular}{lcccccc}
\hline
\hline
Cluster 		& z				& $N_{H}$ $^{a}$ 		& Obs. ID& Exposure Time\\ 
			&				& (cm$^{-2}$)		&	& (ksec)\\\hline
Abell 2204 		& 0.152$^{b}$			& 5.67$\times 10^{20}$	& 7940	& 72.9	\\
\\
Abell 1835		& 0.252$^{b}$			& 2.04$\times 10^{20}$	& 6880	& 110.0	\\
\\
MS~1137.5+6625		& 0.784$^{c}$			& 9.54$\times 10^{19}$	& 536	& 115.5\\
\\
CL~J1226.9+3332		& 0.888$^{d}$			& 1.38$\times 10^{20}$	& 5014	& 32.7\\
			&				&			& 3180	& 31.5\\

\hline  
\hline
\label{table_observations}
\end{tabular}

$(a)$ Leiden/Argentine/Bonn (LAB) Survey, see \citet{kalberla2005}

$(b)$ \citet{struble1999}

$(c)$ \citet{donahue1999}

$(d)$ \citet{ebeling2001}
\end{table*}

\begin{table*}
\centering
\caption{Sources of systematic error in the \chandra\ data \label{tab:systematic}}
\begin{tabular}{cccc}
\hline
\hline
Source of uncertainty & Observable affected & Fractional Error \\
\hline
Background level & Background count rate & 5\% \\
Spatial variations of $A_{eff}$$^{a}$ & Photon count rates & 1\% \\
Energy calibration of $A_{eff}$$^{b}$ & Temperature  measurement & 5\%\\
\hline
\hline
\end{tabular}

$(a)$ Reference: http://cxc.harvard.edu/cal/\\
$(b)$ Reference: http://cxc.harvard.edu/ciao4.1/why/caldb4.1.1\_hrma.html
\end{table*}

\section{Application to \chandra\ Observations of Clusters }
\label{sec_dataAnalysis}

\subsection{Chandra Data Reduction and Analysis}

We use deep \chandra\ ACIS-I observations of
four galaxy clusters to validate our models: two clusters which
do not have a cool core component, MS~1137.5+6625 and CL~J1226.9+3332,
and two cool core clusters, Abell~2204 and Abell~1835.
The observations are summarized in Table \ref{table_observations}. 
As part of the data reduction procedure,
we applied afterglow, bad pixel and charge transfer inefficiency corrections to the 
Level 1 event files using CIAO 4.1 and CALDB 4.1.1. 
Flares in the background due to solar activity are 
eliminated using light curve filtering as described in
\citet{markevitch2003}.
Filtered exposure times 
are also given in Table~\ref{table_observations}. 

For the purpose of background 
subtraction we use blank-sky observations. 
Given that the background is obtained from regions 
of the sky that may have different
soft X-ray fluxes than at the cluster position, we use a peripheral region
of the ACIS-I detector to model the difference between the blank-sky and the
cluster soft fluxes. This step in the analysis is particularly
important for Abell~2204, which lies in a region of significantly higher
soft X-ray emission than the average blank-sky region.
Spectra and images used in this paper are extracted in the energy band 0.7-7.0 keV,
chosen to minimize the effect of calibration uncertainties at the lowest energies,
and the effect of the detector background at high energy.

Spectra are extracted in concentric annuli surrounding the centroid of
X-ray emission after all point sources were removed. An optically 
thin plasma emission model (APEC in XSPEC) is used, with temperature, abundance 
and normalization as free parameters. The redshift and Galactic $N_{H}$ of the 
four clusters are shown in Table 
\ref{table_observations}. 


\subsection{Systematic Uncertainties in the \chandra\ Data Analysis}

We consider possible sources of systematic uncertainty in the \chandra\ data.
The blank-sky background  used in our analysis is normalized
to the high-energy background level of each cluster observation,
determined from peripheral regions
of the ACIS detector that are free of cluster emission
\citep[following][]{markevitch2003}.
The primary source of uncertainty in the background subtraction
is the choice of a peripheral region as representative
of the background at the cluster location. Due to the
scatter in the count rate of various peripheral regions
in each cluster observation,
we estimate a $\sim$5\% uncertainty
in the determination of the background level
from these \chandra\ observations. We use this uncertainty
in the spectral and imaging data analysis.

\begin{figure*}
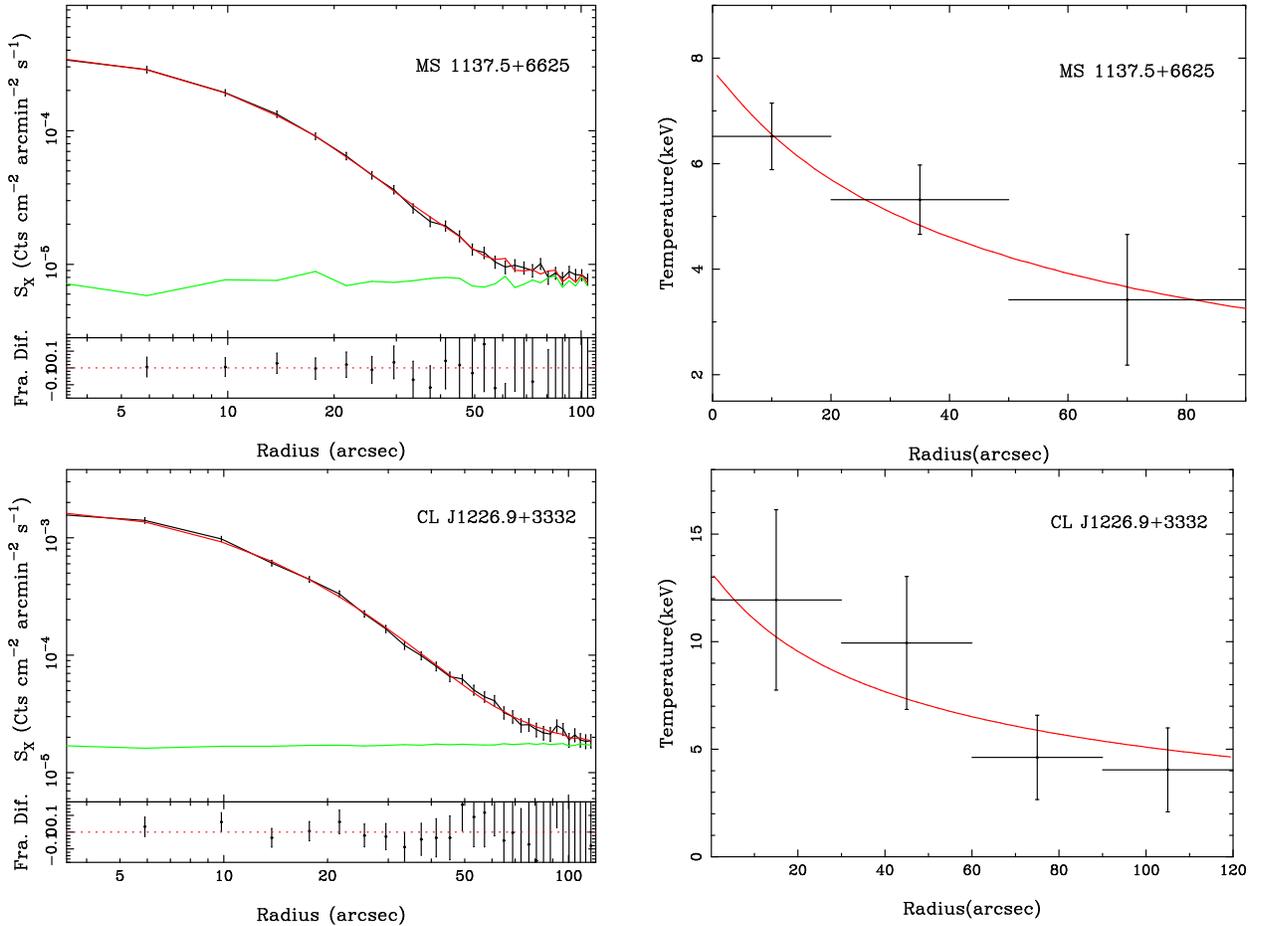

\centering
\resizebox{\textwidth}{!}{
\includegraphics[angle=-90,width=1cm]{f5a.eps}
\includegraphics[angle=-90,width=1cm]{f5b.eps}
}
\resizebox{\textwidth}{!}{
\includegraphics[angle=-90,width=1cm]{f6a.eps}
\includegraphics[angle=-90,width=1cm]{f6b.eps}
}
\caption{X-ray surface brightness and temperature profiles of MS~1137.5+6625
in the radial range 0 - 90 \arcsec and CL~J1226.9+3332 in the radial range 0 - 120 \arcsec.
The red line in both profiles 
shows the best fit model to the data; the green line in surface brightness 
profiles shows the background level. The overall $\chi^{2}$ of the fit (Table~\ref{table_bestFitParams})
is the sum of the $\chi^{2}$ values of the surface brightness profile and the
temperature profile.}
\label{fig:ms1137-cl1226}
\end{figure*}

Calibration of the ACIS effective area is another significant source
of systematic uncertainty in our analysis.
For the spectral data used for measuring the gas temperature,
the primary source of uncertainty is the low-energy calibration
of the effective area and the presence of a contaminant on
the optical filter of the ACIS detector.
We use the \chandra\ calibration available in CALDB 4.1.1, which includes
 a significant change in the effective area  calibration
which improves the agreement between clusters temperatures 
obtained with ACIS-I, and also with other instruments (such as \xmm's EPIC). 
With this calibration of the \chandra\ efficiency we estimate that
any residual systematic error in the measurement of cluster temperatures
is of order $\sim$5\%, and add this error to the temperature measured in 
each bin.

For the imaging data,
spatially-dependent non-uniformities
in the ACIS efficiency are a relevant source
of possible systematic error because of the extended
nature of the sources.
The absolute calibration of the ACIS efficiency is currently at
the level of 3\%, with possible spatial variations on arcmin scales
at the level of $\sim$1\%; we use a 1\% error as additional
uncertainty in the count rates for each annulus.

Table~\ref{tab:systematic} provides a summary of the
uncertainties included in our analysis of the \chandra\ data,
and references to the \chandra\ calibration information.


\subsection{Result of Model Fits and Mass Measurements}
\label{sec_massMeasurements}

The radial profiles of the X-ray surface brightness 
and temperature observed from the
\chandra\ data 
are used to determine the best-fit parameters and the goodness
of fit for MS~1137.5+6625, CL~J1226.9+3332, Abell~1835 and Abell~2204.
The X-ray surface brightness is

\begin{equation}
S_{x}=\frac{1}{4\pi (1+z)^{3}}\int  n_{e}^{2} \Lambda_{ee}(T) \, dl, 
\label{eqn:surface_brightness}
\end{equation}

where $S_{x}$ is in detector units
(counts cm$^{-2}$ arcmin$^{-2}$ s$^{-1}$), $z$ is the cluster redshift,
$\Lambda_{ee}(T)$ is the plasma emissivity in detector units
(counts cm$^{3}$ s$^{-1}$) which we calculate
using the APEC code \citep{smith2001}
and $l$ is the distance along the line of sight.

We validate the model using two polytropic clusters which do not have
a cool core component, MS~1137.5+6625 
and CL~J1226.9+3332, and two cool core clusters Abell~1835 and Abell~2204.
We use the Monte Carlo Markov Chain code described in \citet{bonamente2004}
for the fit.
The model described in \S~\ref{sec_models} has 8 free parameters,
of which 5 parameters describe the global cluster properties
($n_{e0},\ T_{e0},\ r_{s},\ \beta,\ n$)
and 3 additional parameters ($r_{cool},\ \alpha, \ \gamma$) are used to 
model the central region of the cool core clusters.

\begin{table*}
\begin{center}
\tiny
\caption{Best-fit parameters of the model}
\begin{tabular}{lcccccccccc}
\hline\hline
Cluster         & $n_{e0}$              & $r_{s}$               & $n$           & $\beta$       &
$T_{0}$         & $r_{cool}$            &$\alpha$              & $\gamma$        & $\chi^{2}$ (d.o.f.) & \textit{P} value  \\
                & ($10^{-2} cm ^{-3}$)& (\textit{arcsec}) &       &       &
($keV$)       	& (\textit{arcsec})       &                       &               &  & \\ \hline
\\

MS~1137.5+6625	& 2.17$^{+0.07}_{-0.19}$ & 23.84$^{+10.88}_{-2.19}$ & 4.83$^{+1.30}_{-0.27}$
& 2.0             & 7.84$^{+1.08}_{-1.26}$ &-& -
&-                & 14.4 (26) & 96.7 \%                \\
\\
CL~J1226.9+3332& 3.99$^{+0.23}_{-0.23}$ & 22.65$^{+5.06}_{-2.79}$ & 4.54$^{+0.54}_{-0.31}$
& 2.0		& 13.32$^{+2.28}_{-2.98}$ &-& -
&-                & 14.6 (29)& 98.8\%                \\
\\
Abell 2204      & 4.42$^{+0.37}_{-0.24}$ & 21.73$^{+1.50}_{-2.01}$ & 6.44$^{+1.02}_{-0.51}$
& 1.39$^{+0.04}_{-0.06}$                 & 14.28$^{+0.75}_{-0.78}$ & 19.42$^{+0.60}_{-0.73}$  & 0.16$^{+0.01}_{-0.01}$
& 2.0           & 115.5 (145)  & 96.6\%                \\
\\
Abell 1835      & 2.57$^{+0.29}_{-0.07}$ & 40.32$^{+3.29}_{-6.48}$ & 3.98$^{+0.73}_{-0.41}$
& 1.94$^{+0.15}_{-0.22}$                 & 18.26$^{+0.42}_{-1.61}$ & 22.65$^{+0.28}_{-1.17}$  & 0.18$^{+0.02}_{-0.01}$
& 2.0	        & 99.3 (93)     	& 30.8\%        \\
\\
\hline  \hline
\label{table_bestFitParams}
\end{tabular}
\end{center}
\end{table*}

\begin{figure*}
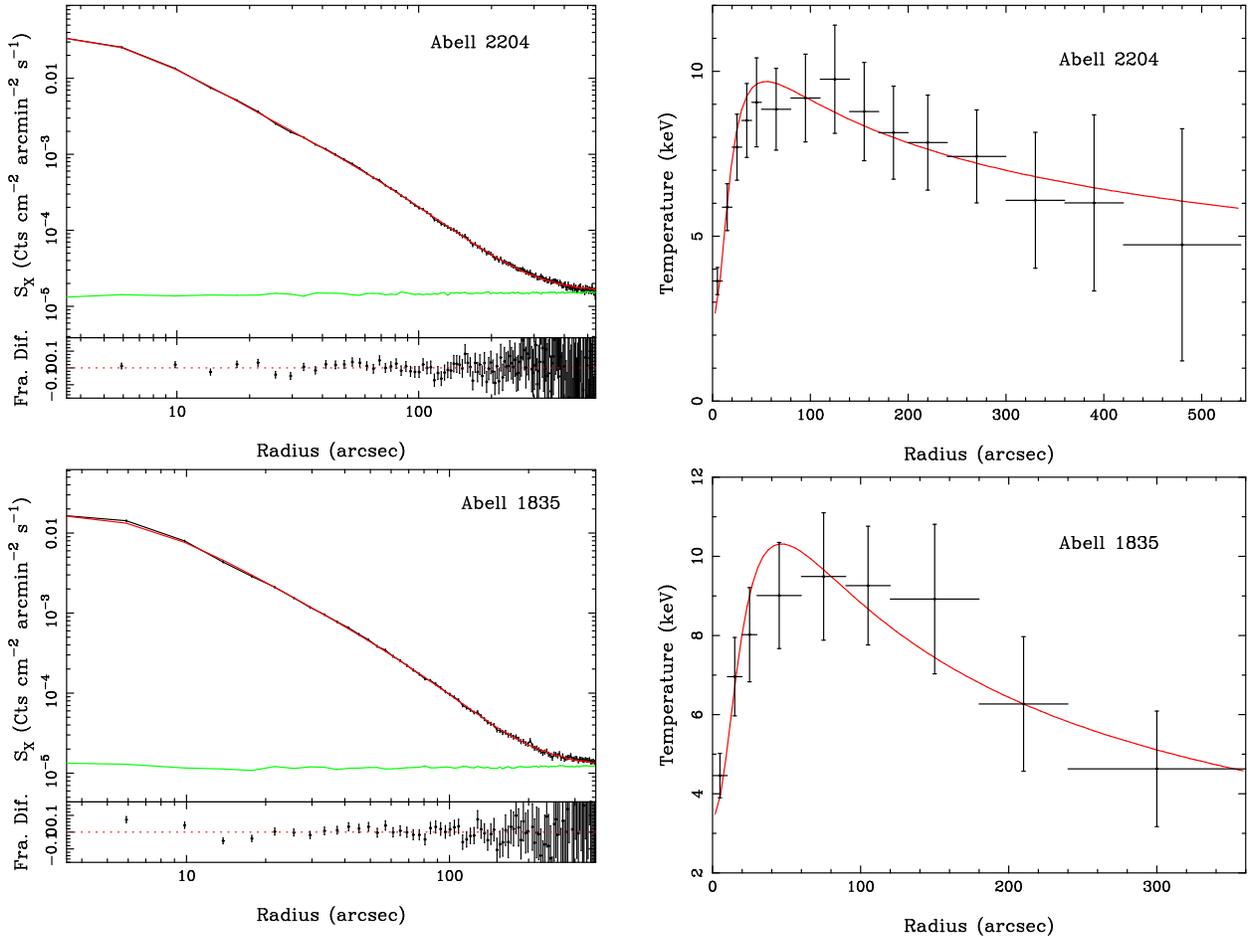

\centering
\resizebox{\textwidth}{!}{
\includegraphics[angle=-90,width=1cm]{f3a.eps}
\includegraphics[angle=-90,width=1cm]{f3b.eps}
}
\resizebox{\textwidth}{!}{
\includegraphics[angle=-90,width=1cm]{f4a.eps}
\includegraphics[angle=-90,width=1cm]{f4b.eps}
}
\caption{X-ray surface brightness and temperature profiles of cool core
clusters Abell~2204 in the radial range 0 - 540 \arcsec and Abell~1835 
in the radial range 0 - 360 \arcsec. The red line in both profiles 
shows the best fit model to the data; the green line shows 
the background level. The overall $\chi^{2}$ of the fit (Table~\ref{table_bestFitParams})
is the sum of the $\chi^{2}$ values of the surface brightness profile and the
temperature profile.}
\label{fig:a2204-a1835}
\end{figure*}


\subsubsection{Polytropic Clusters MS~1137.5+6625 and CL~J1226.9+3332}

For polytropic clusters, which do not have a cool core component,
5 parameters are sufficient to describe the distribution of
density, temperature.
We fixed $\beta$ to 2 for these clusters \citep{navarro1996}, since the 
polytropic index (\textit{n}) and $\beta$
cannot both be determined from X-ray data available.
We report the $\chi^{2}$ of the best-fit 
model for MS~1137.5+6625 and CL~J1226.9+3332 in Table~\ref{table_bestFitParams}.
We also calculate the gas mass by taking the volume integral of Equation \ref{eqn:ne0_poly}
and the total mass using Equation \ref{eqn_gen_nfw_totalMass} 
and report the results in Table~\ref{table_mass}.

\begin{table*}
\centering
\caption{Gas and total masses}
\begin{tabular}{@{\extracolsep{\fill}}lcccccc}
\hline\hline
Cluster         & $r_{2500}$            &$M_{gas}(r_{2500})$    	& $M_{tot}(r_{2500})$   & $r_{500}$     &$M_{gas}(r_{500})$ & $M_{tot}(r_{500})$ \\
                &(arcsec)               & ($10^{13} M_{\odot}$) 	& ($10^{14} M_{\odot}$) & (arcsec)      & ($10^{13} M_{\odot}$) & ($10^{14} M_{\odot}$)  \\
\hline
\\
MS~1137.5+6625  & 44.7$^{+4.1}_{-3.8}$ & 1.10$^{+0.13}_{-0.11}$		& 1.20$^{+0.36}_{-0.28}$
                & 98.5$^{+9.6}_{-8.2}$ & 3.08$^{+0.13}_{-0.13}$       	& 2.56$^{+0.82}_{-0.59}$ \\
\\ 

CL~J1226.9+3332 & 48.8$^{+4.7}_{-5.3}$ & 3.01$^{+0.39}_{-0.44}$		& 2.16$^{+0.69}_{-0.63}$
                & 104.6$^{+9.1}_{-10.4}$& 8.29$^{+0.55}_{-0.63}$       & 4.25$^{+1.22}_{-1.14}$ \\
\\
Abell~2204      & 225.7$^{+4.1}_{-4.1}$ & 3.99$^{+0.09}_{-0.09}$ 	& 3.37$^{+0.19}_{-0.18}$
                & 479.8$^{+11.4}_{-11.2}$ & 10.35$^{+0.26}_{-0.26}$	& 6.47$^{+0.47}_{-0.44}$ \\
\\
Abell~1835      & 150.6$^{+3.4}_{-4.2}$ & 4.97$^{+0.14}_{-0.17}$	& 3.72$^{+0.26}_{-0.30}$
                & 309.7$^{+9.8}_{-13.1}$ & 12.08$^{+0.38}_{-0.50}$     & 6.47$^{+0.64}_{-0.79}$ \\
\\
\hline  \hline
\label{table_mass}
\end{tabular}
\end{table*}


\subsubsection{Cool Core Clusters Abell~2204 and Abell~1835}

The clusters Abell 2204 and Abell 1835 have a clear 
cool core component (see Figure \ref{fig:a2204-a1835}),
which requires the use of the cooling equations described in \S
\ref{sec_models_coolcore}. The best fit model parameters 
are listed in Table~\ref{table_bestFitParams}. 
We also calculate the gas mass by taking the volume
integral of Equation \ref{eqn_coolcore_gas_density} 
and the total mass using Equation \ref{eqn:gen_nfw_totalMass2} 
and report the results in Table~\ref{table_mass}.

\begin{table*}
\centering
\scriptsize
\caption{Mass Comparison of Abell~1835 
with \citet{mroczkowski2009}}
\begin{tabular}{lcccccc}
\hline\hline
	& $r_{2500}$		& $M_{2500,gas}$		& $M_{2500,tot}$	
&$r_{500}$		& $M_{500,gas}$		& $M_{500,tot}$	\\
			& arcsec		& $10^{13}\ M_{sun}$		& $10^{14}\ M_{sun}$	
& arcsec		& $10^{13}\ M_{sun}$	& $10^{14}\ M_{sun}$	\\\hline
\\
\textbf{Abell 1835}	&&& &&&\\
\\
\citet{mroczkowski2009}& 169.0$^{+5.5}_{-8.0}$		& 5.77$^{+0.25}_{-0.35}$	& 5.30$^{+0.53}_{-0.72}$ [Note(a)]
			&363.0 $^{+17.0}_{-12.0}$	& 13.94$^{+0.64}_{-0.52}$	& 10.68$^{+1.54}_{-1.01}$ [Note(b)]\\
\\
Polytropic Model (this work)& 169.0$^{+7.0}_{-7.0}$	& 5.79$^{+0.35}_{-0.46}$	& 4.13$^{+0.31}_{-0.32}$
			   &363.0 $^{+15.0}_{-15.0}$	& 14.31$^{+0.71}_{-0.82}$	& 7.37$^{+0.82}_{-0.83}$\\
\\\hline  \hline
\label{table_massComparison}
\end{tabular}

$(a)$ When X-ray temperature is recalibrated using CALDB 4.1.1, mean value of $M_{2500,tot}$ decreases to $\sim 4.40 \times 10^{14}\ M_{sun}$ \\
$(b)$ When X-ray temperature is recalibrated using CALDB 4.1.1, mean value of $M_{500,tot}$ decreases to $\sim 8.86\times 10^{14}\ M_{sun}$ \\
\end{table*}


\section{Comparison with Previous Work }
\label{sec:comparison}

In Table \ref{table_massComparison} we present the comparison of mass measurements  
of Abell 1835
produced from the polytropic model with the masses reported in \citet{mroczkowski2009}
at $r_{2500}$ and $r_{500}$.
For this purpose we calculate gas mass and total mass at the same radii 
$r_{2500}$ and $r_{500}$ as in \citet{mroczkowski2009}, and use the same
Gaussian uncertainty on $r_{\Delta}$ in order to have a fair comparison on masses.
The gas mass measurements produced by the polytropic model  
are consistent with the \citet{mroczkowski2009} results at the 1$\sigma$
level (see Table \ref{table_massComparison}).

The Chandra Calibration Database (CALDB) has recently been revised
to correct the effective area, resulting in lower X-ray temperatures,
especially for massive clusters \footnote[1]{$http://cxc.harvard.edu/caldb/downloads/Release\_notes/\\
CALDB\_v4.1.1.html$}. The peak X-ray temperatures reported by 
\citet{mroczkowski2009} using CALDB 3.4 are $\sim$ 2 keV 
greater than the temperatures derived in this paper using 
the recent calibration (CALDB 4.1.1). From Equations \ref{eqn_gen_nfw_totalMass}
and \ref{eqn:gen_nfw_totalMass2}, we estimate that this temperature
change would reduce the total masses reported by \citet{mroczkowski2009}
by 17 \% (see Table \ref{table_massComparison}). When the X-ray
temperature calibration  issue is accounted for, 
the total mass values in Table \ref{table_massComparison} are
in agreement within the stated 1$\sigma$ uncertainties.


\section{Discussion and Conclusions}
\label{sec_conclusion}

We introduce a new model to describe 
the physical properties 
of the hot intra-cluster medium and present an application of the model to
\chandra\ X-ray observations of MS~1137.5+6625, CL~J1226.9+3332, 
Abell~1835 and Abell~2204. The model is based on a 
polytropic equation of state for the gas in hydrostatic
equilibrium with the cluster gravitational potential.
Using a function for the cluster total mass density that has the
asymptotic slope as a free parameter,
we obtain analytic expressions for the gas density, temperature and
pressure. We also include a core taper function 
that accounts for the cooling of the gas in the cluster center.

This model has a number of features that make it
suitable for the analysis of
X-ray and SZE observations of galaxy clusters.
The model is analytic, and has a limited number of parameters which
describe the global properties of the cluster. For clusters 
which do not have a cool core,
5 parameters are sufficient to describe the distribution of
density, temperature, pressure and total matter density.
The gas density and temperature are linked by the polytropic equation
of state, and the total matter density is related to the plasma
properties by the hydrostatic equation. Therefore there is
just one scale radius ($r_s$) that appears in the radial distribution
of all thermodynamic quantities. The other parameters that describe the global
physical properties of the cluster are the central density ($n_{e0}$) and temperature
($T_{0}$) of the gas, the polytropic index $n$, and the asymptotic slope
of the total mass density ($\beta+1$).
For cool core clusters, three additional parameters allow an accurate
description of the cooling of the gas in the core, and the accompanying
increase in the density (\S~\ref{sec_models_coolcore}). 

In addition to the analysis of spatially-resolved spectroscopic
and imaging X-ray data (see \S \ref{sec_dataAnalysis}), the model is applicable to 
SZE observations, which require a model for the plasma pressure.
A number of models suitable for SZE observations are available in the
literature, for example \citet{nagai2007} and \citet{mroczkowski2009}.
Our model has the advantage of the simultaneous applicability to
both X-ray and SZE observations, 
and it is therefore suitable for a number of cosmological
applications including the measurement of
the Hubble constant \citep[][]{bonamente2006},
the measurement of scaling relations between X-ray and
SZE observables \citep{bonamente2008}, the measurement
of cluster masses independent of cosmology
from joint X-ray and SZE data (Hasler et al. 2010) and the measurement of
the effect of \he sedimentation on X-ray measured masses \citep{bulbul2010}.

\section*{Acknowledgments} The authors would like to thank the referee, J. Carlstrom, 
D. Marrone and T. Mroczkowski for their useful comments on the manuscript.

\end{document}